\documentclass[aps,pre,superscriptaddress]{revtex4-1}
\usepackage{amsmath,amssymb,latexsym,epsfig,graphics,epsf}
\usepackage{graphics}
\usepackage{float}
\newcommand{\fig}[1]{Fig.~\ref{#1}}
\newcommand{\eq}[1]{Eq.~(\ref{#1})}

\newcommand{\e}{{\rm EXP}}

\newcommand{\bR}{\bold R}

\newcommand{\be}{\begin{equation}}
\newcommand{\ee}{\end{equation}}

\newcommand{\Sex}{{S}_{\rm ex}}

\begin{document}
\title{Studies of the Lennard-Jones fluid in 2, 3, and 4 dimensions highlight the need for a liquid-state $1/d$ expansion}
\author{Lorenzo Costigliola, Thomas B. Schr{\o}der, and Jeppe C. Dyre}
\affiliation{"Glass and Time", IMFUFA, Dept. of Science and Environment, Roskilde University, P. O. Box 260, DK-4000 Roskilde, Denmark}

\date{\today}

\begin{abstract}
The recent theoretical prediction by Maimbourg and Kurchan [arXiv:1603.05023] that for regular pair-potential systems the virial potential-energy correlation coefficient increases towards unity as the dimension $d$ goes to infinity is investigated for the standard 12-6 Lennard-Jones fluid. This is done by computer simulations for $d=2,3,4$ going from the critical point along the critical isotherm/isochore to higher density/temperature. In all cases the virial potential-energy correlation coefficient increases significantly. For a given density and temperature relative to the critical point, with increasing number of dimension the Lennard-Jones system conforms better to the hidden-scale-invariance property characterized by high virial potential-energy correlations (a property that leads to the existence of isomorphs in the thermodynamic phase diagram, implying that it becomes effectively one-dimensional in regard to structure and dynamics). The present paper also gives the first numerical demonstration of isomorph invariance of structure and dynamics in four dimensions. Our findings emphasize the need for a universally applicable $1/d$ expansion in liquid-state theory; we conjecture that the systems known to obey hidden scale invariance in three dimensions are those for which the yet-to-be-developed $1/d$ expansion converges rapidly.
\end{abstract}
\maketitle

Recent years have brought notable progress in the understanding of the liquid state coming from studies of the high-dimensional limit. With roots back in time \cite{ree64,fri87,lub90,fri93,fri99} 
and in a continuation of recent progress \cite{par99,par00,par08,cha11a,kur12},  
Charbonneau and collaborators in 2014 in a {\it tour de force} replica symmetry breaking calculation solved the glass problem in high dimensions for the prototypical hard-sphere (HS) model \cite{cha14}. This was followed by a proof by Maimbourg, Kurchan, and Zamponi that the dynamics satisfies a universal equation in high dimensions for the general case of a system of particles interacting via pairwise additive forces \cite{mai16a}. This is how a ``simple'' liquid is traditionally defined  \cite{ric65,tem68,barrat,kir07,han13}, although during the last 20 years it has gradually become clear that some such systems -- like the Gaussian core model, the Lennard-Jones Gaussian model, and the Jagla model -- exhibit quite complex behavior (see, e.g., Ref. \onlinecite{ing12} and its references).

Very recently, Maimbourg and Kurchan showed that in the condensed phase, i.e., for states dominated by hard repulsions, any well-behaved pair-potential system has strong virial potential-energy correlations in sufficiently high dimensions \cite{mai16,kur16}. Specifically, it was shown that the Pearson correlation coefficient $R$ of the constant-volume canonical-ensemble equilibrium fluctuations of virial $W$ and potential energy $U$,

\be\label{R}
R=\frac{\langle\Delta W\Delta U\rangle}{\sqrt{\langle(\Delta W)^2\rangle\langle(\Delta U)^2\rangle}}\,,
\ee
converges to unity as the number of dimensions $d$ goes to infinity. The analysis presented in Ref. \onlinecite{mai16} also showed that the $\e$ pair potential (a simple exponential decay in space) plays the role as a building block of all pair potentials \cite{bac14,bac14a}. Note that, in contrast to the inverse-power-law pair potentials $\propto r^{-n}$ ($r$ being the pair distance), due to its rapid spatial decay the $\e$ pair potential has a thermodynamic limit in all dimensions.  

Systems with $R$ close to unity are characterized by ``hidden scale invariance'', an approximate symmetry that has been studied in several publications since its introduction in 2008; there are now also experimental verifications of the concept for van der Waals liquids \cite{gun11,boh12,roe13,xia15}. Systems with hidden scale invariance are simple because they have so-called isomorphs in the thermodynamic phase diagram, which are lines along which structure and dynamics in suitably reduced units are invariant to a good approximation. The isomorph theory has been applied to atomic and molecular liquid and crystalline models in thermal equilibrium, as well as to non-equilibrium phenomena like shear flows of liquids and glasses (see, e.g., Ref. \onlinecite{dyr14} and its references). Recently, it was shown from state-of-the-art DFT {\it ab initio} simulations of 58 liquid elements at their triple points that most metals possess hidden scale invariance \cite{hum15}. An overview of the isomorph theory was given in Ref. \onlinecite{dyr14} from 2014. After that paper was written, it became clear that Roskilde (R) simple systems \cite{mal13,abr14,fer14,fle14,pra14,buc15,grz15,har15,hey15,ing15,kip15,sch15,khr16,adr16} -- those with $R>0.9$ -- are characterized by approximately obeying the following condition \cite{sch14}: $U(\bR_a)=U(\bR_b)\Leftrightarrow U(\lambda\bR_a)= U(\lambda\bR_b)$ in which $\bR$ specifies all particle positions and $U(\bR)$ is the potential-energy function. Thus hidden scale invariance is equivalent to an approximate conformal invariance property. 

The non-trivial finding of the above-mentioned works is that many realistic model systems -- as well as many real-world liquids and solids --  obey hidden scale invariance. It appears that most metals and van der Waals bonded liquids and solids exhibit hidden scale invariance, whereas systems with strong directional bonds like covalently or hydrogen-bonded systems do not and are generally more complex \cite{dyr14}.

This paper presents computer simulations of the standard 12-6 Lennard-Jones (LJ) system in two, three, and four dimensions consisting of particles interacting via the pair potential 

\be\label{LJ}
v_{\rm LJ}(r)
\,=\,4\varepsilon\left[ \left(\frac{r}{\sigma}\right)^{-12} - \left(\frac{r}{\sigma}\right)^{-6} \right] \,.
\ee
Here $\varepsilon$ and $\sigma$ define the characteristic energy and length scales of the pair potential. The LJ system does not have a proper thermodynamic limit in more than five dimensions, and one may argue what is the correct generalization of this system to arbitrary dimension $d$ (for instance, $v(r)\propto (r/\sigma)^{-(d+9)}-(r/\sigma)^{-(d+3)}$ or $v(r)\propto (r/\sigma)^{-4d}-(r/\sigma)^{-2d}$ or a third option). We avoided this problem by staying at low dimensions. 

{It is not obvious how to compare results for different thermodynamic state points in different dimensions. In high dimensions one may compare different state points by scaling the density such that the HS packing fraction remains invariant \cite{bis07}. In our case of relatively low dimensions this is too crude; in any case, we also need a scaling of the temperature in order to be able to compare results obtained in different dimensions. The critical point of the LJ system is known for $d=2,3,4$ \cite{smi91,oku00,hlo99}, and we used this as reference state point. This choice has the further advantage that in three dimensions the virial potential-energy correlations are weak in the vicinity of the critical point, which allows one to monitor how $R$ increases when the condensed ``strongly correlating'' liquid phase is approached upon increasing density or temperature.

Molecular dynamics simulations have been performed before in four spatial dimensions \cite{van93,bis99,sco99}. The simulations reported below used a homemade code applicable in arbitrary dimensions \cite{cos16a}. The code implements $NVT$ dynamics with periodic boundary conditions \cite{tildesley} based on the leap-frog algorithm coupled with a Nose-Hoover thermostat. A shifted-forces cutoff at $2.5 \sigma$ was used in all simulations \cite{tox11a}. The time step $t_s$ varied with state point such that the reduced time step, $\tilde{t}_s\equiv t_s \rho ^{1/d} \sqrt{k_B T /m}$, was 0.001 (here $\rho\equiv N/V$ is the particle density and $m$ the particle mass). After melting and equilibrating from a simple cubic configuration, the LJ system was simulated at every liquid state point for $2\cdot 10^7$ time steps. In two dimensions the system crystallized at the three highest-density state points; simulations at these state points were performed with a reduced time step of $\tilde{t}_s=0.0005$ and the number of time steps was doubled. In all cases the thermostat relaxation time was $80$ time steps. The system size was $N=1225$ in two, $N=1728$ in three, and $N=2401$ in four dimensions.

\begin{figure}
	\begin{center}
		\includegraphics[width=7cm]{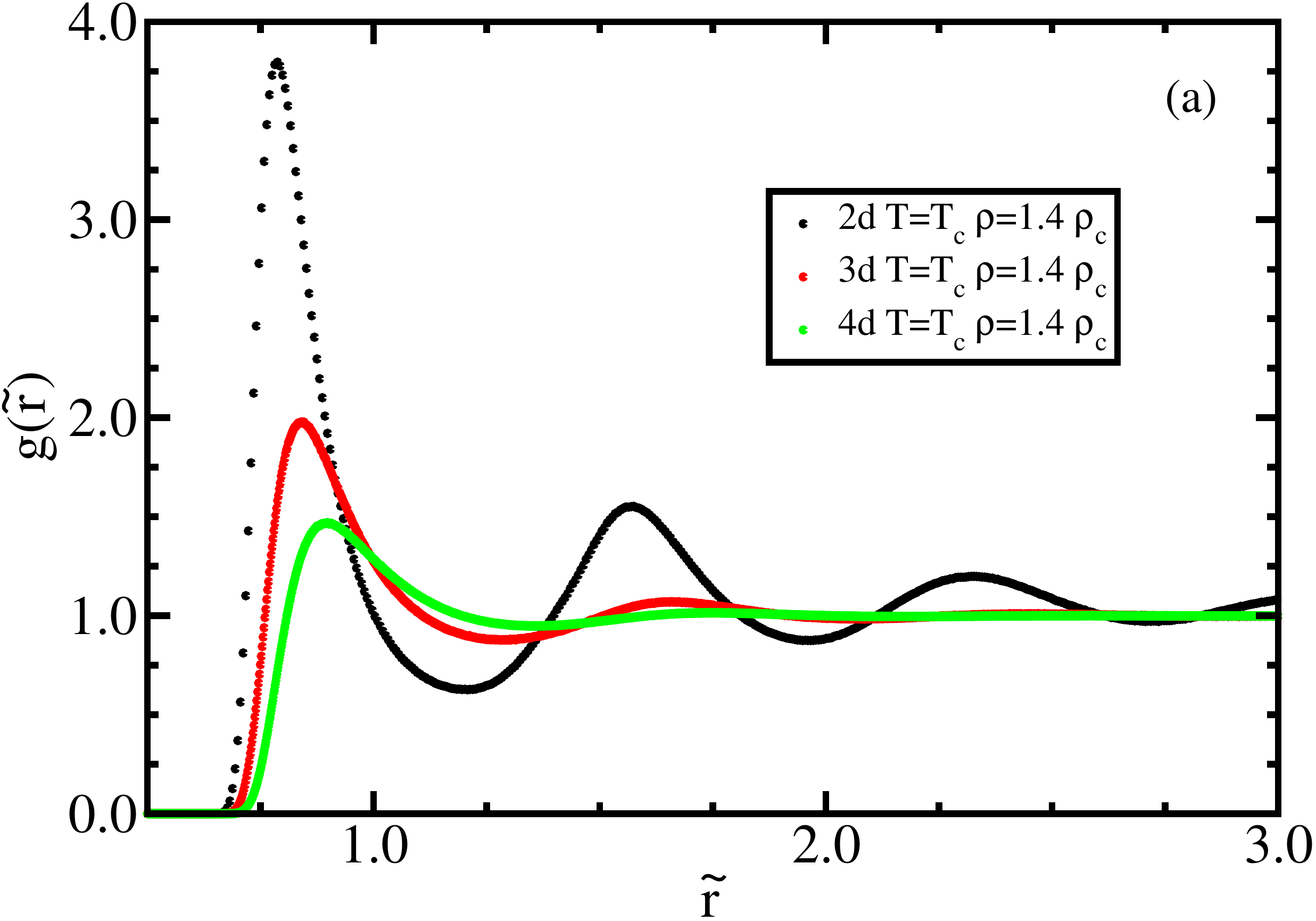} 
		\includegraphics[width=7cm]{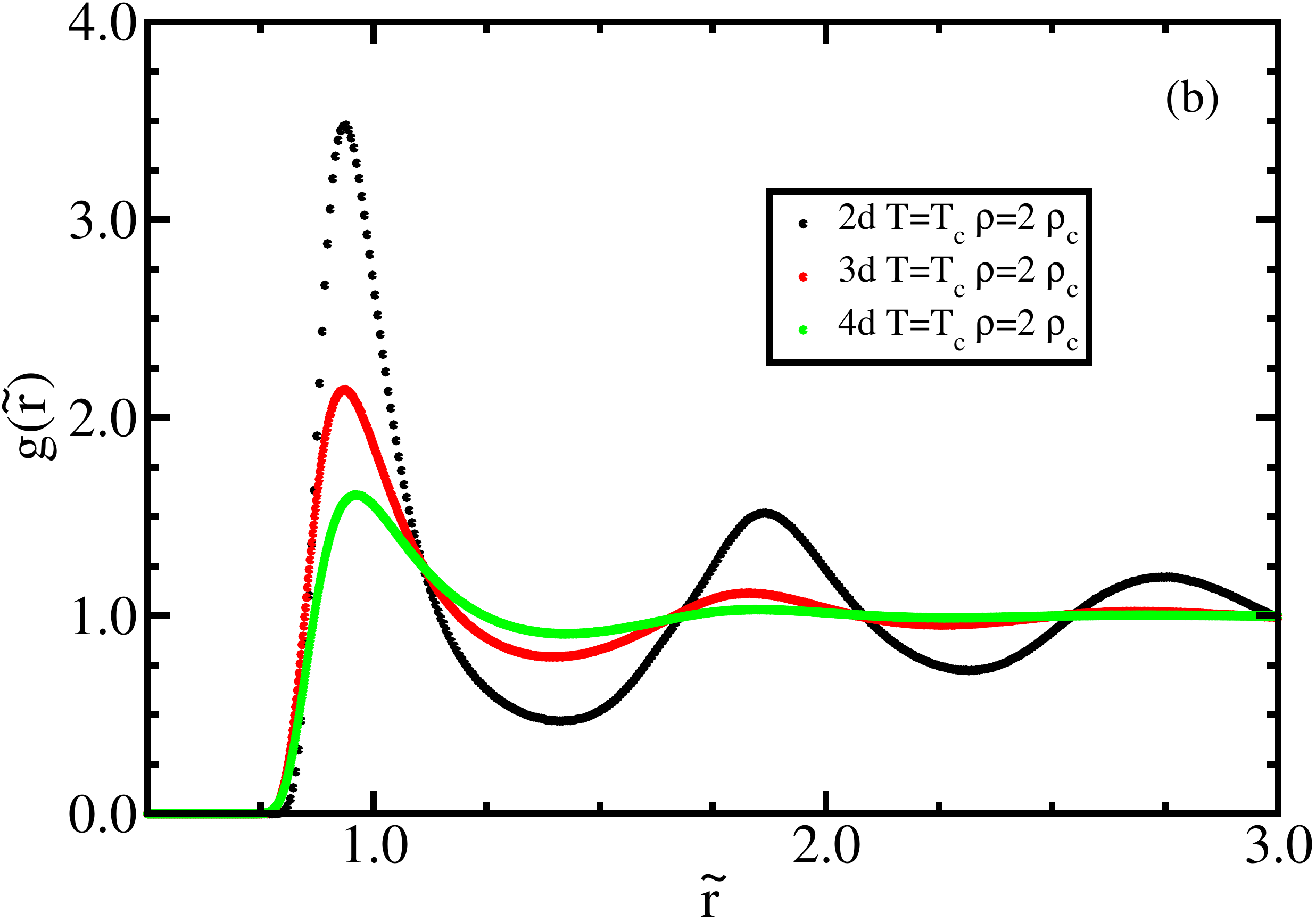} 
	\end{center}
	\caption{Radial distribution function of the Lennard-Jones (LJ) fluid along the critical isotherm in two, three, and four dimensions (black, red, and green colors, respectively; $\tilde r\equiv \rho^{1/d}r$ where $\rho$ is the particle density and $r$ the interparticle distance). 
		(a) shows results at $1.4$ times the critical density $\rho_c$, (b) shows results at twice the critical density. In both cases the fluid's long-range structure becomes markedly less pronounced as the number of dimensions increases. 
		\label{str_fig}}
\end{figure}

Our focus is on what happens in the fluid region of phase space in which the correlation coefficient $R$ of \eq{R} is far from unity in 3d. This number {\it is} close to unity in the ``ordinary'' 3d condensed liquid phase not too far from the melting line, as well as in the entire crystalline phase \cite{I,bai14,alb14}, but approaching the gas phase and, in particular, the critical point in 3d, $R$ drops quickly and the system is no more R simple \cite{bai14}. 

Figure \ref{str_fig} reports the reduced-unit radial distribution function $g(r)$ at the critical temperature $T_c$ at $1.4$ and $2$ times the critical density; the black symbols mark $g(r)$ in two dimensions, the red curves in three dimensions, and the green curves in four dimensions. Figure \ref{str_fig} nicely confirms the argument of Maimbourg and Kurchan that in higher dimensions the nearest-neighbor distance increasingly dominates the physics \cite{mai16}. Thus beyond the first coordination shell $g(r)$ converges quickly to unity in high dimensions. In the words of Ref. \onlinecite{mai16}, what happens in high dimensions is that a single pair distance dominates the physics because ``particles that are too close are exponentially few in numbers, while those that are too far interact exponentially weakly''. This argument presupposes, of course, that the pair potential in question has been generalized to any number of dimensions in a way ensuring a proper thermodynamic limit, i.e., such that it decays more rapidly than $r^{-d}$ at long distances. 

A system for which a single pair distance dominates the physics even in three dimensions is the hard-sphere (HS) system for which the radial distribution function at contact determines the equation of state \cite{son88,son89}. The above suggests that one may regard the $3d$ HS system as {\it a poor man's version of the $d\rightarrow\infty$ limit;} indeed, it has been known for some time that the pair correlations of the HS system becomes increasingly trivial as $d$ increases \cite{bis05,bis07,whi07}. We note, however, that the HS system is not the only possibility of a $3d$ poor man's $d\rightarrow\infty$ limit; alternatives are the Gaussian core model \cite{cos16b} or the Mari-Kurchan model \cite{mar11a}.

\begin{figure}
	\begin{center}
		\includegraphics[width=7cm]{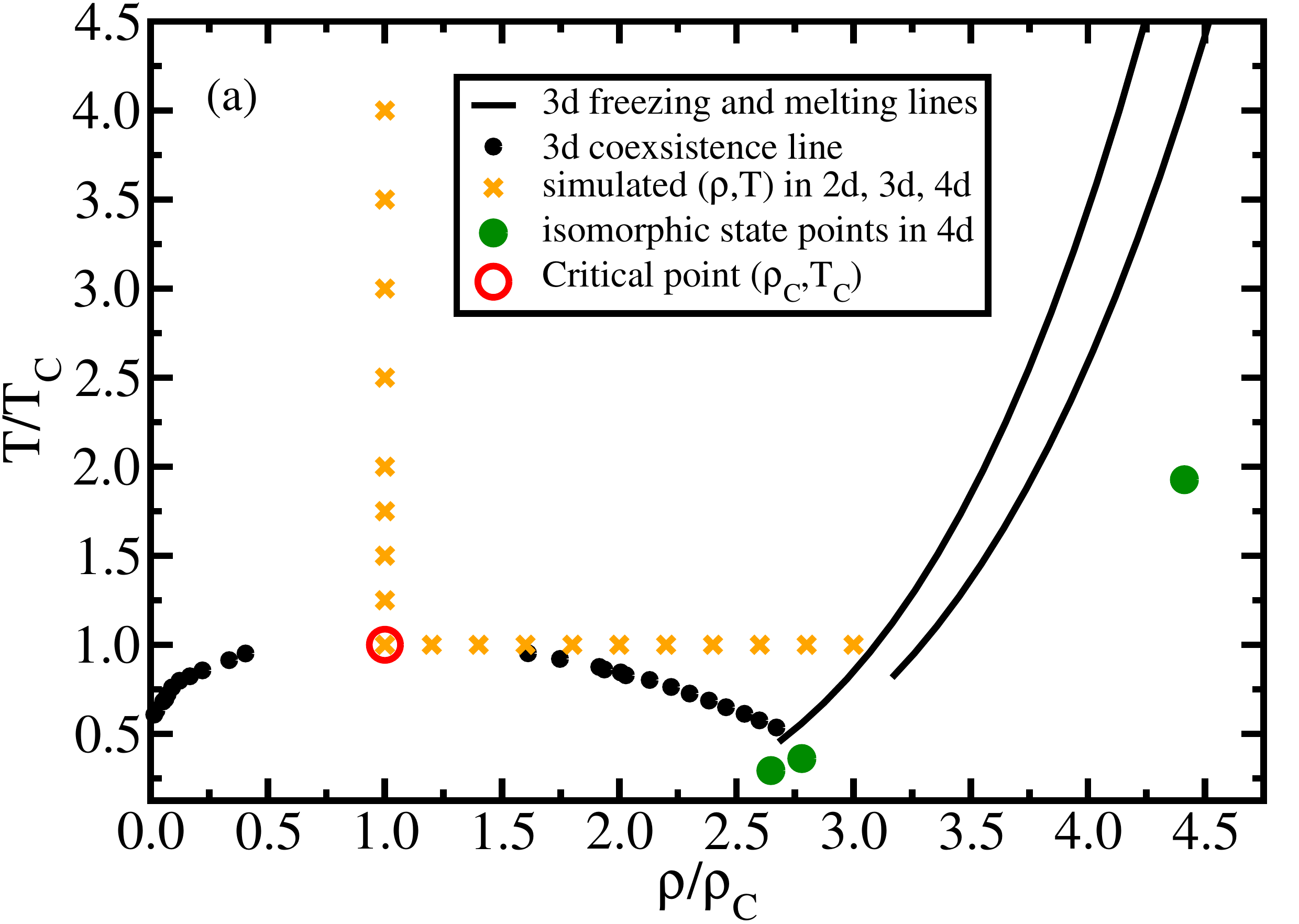} 
		\includegraphics[width=7cm]{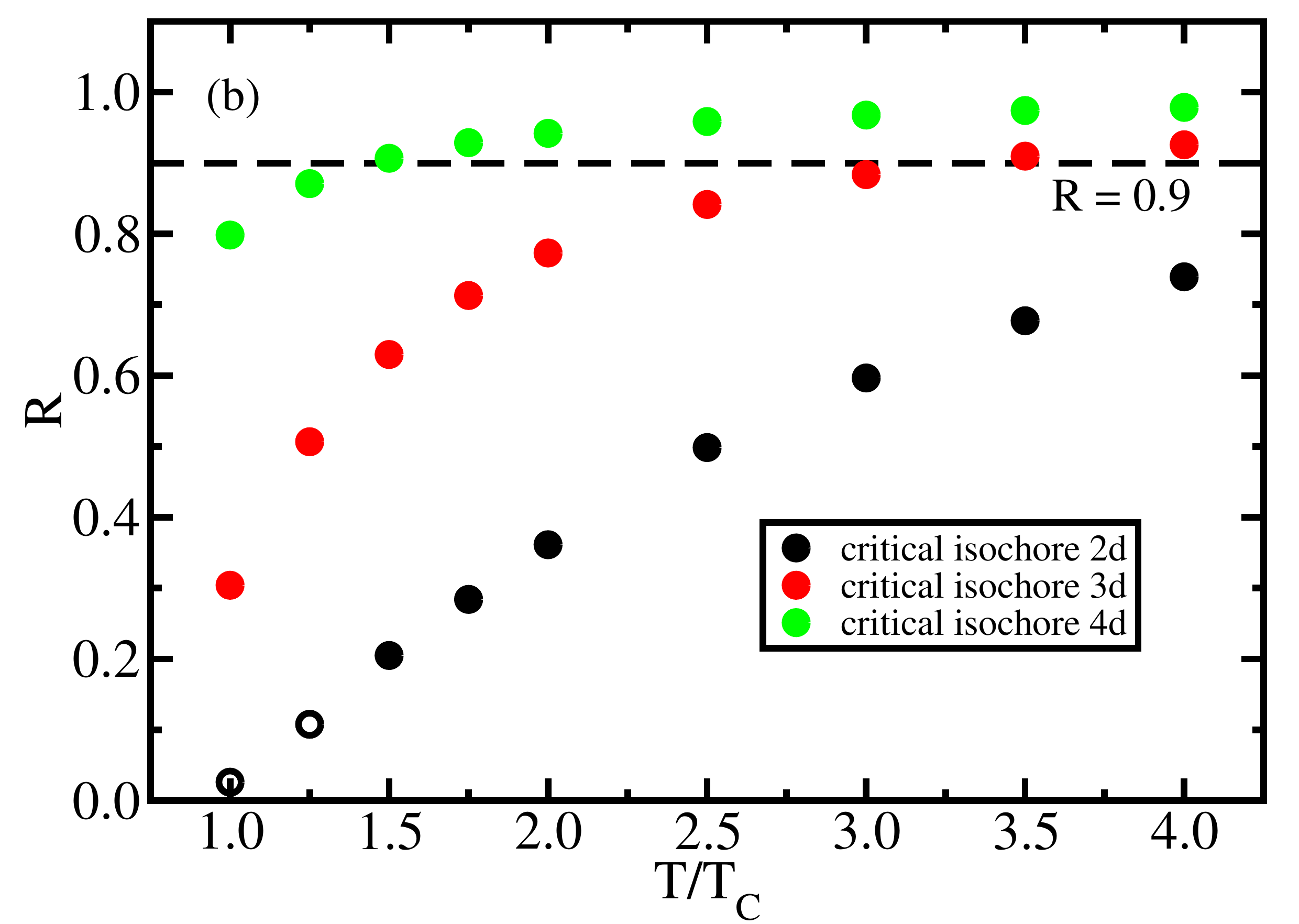} 
		\includegraphics[width=7cm]{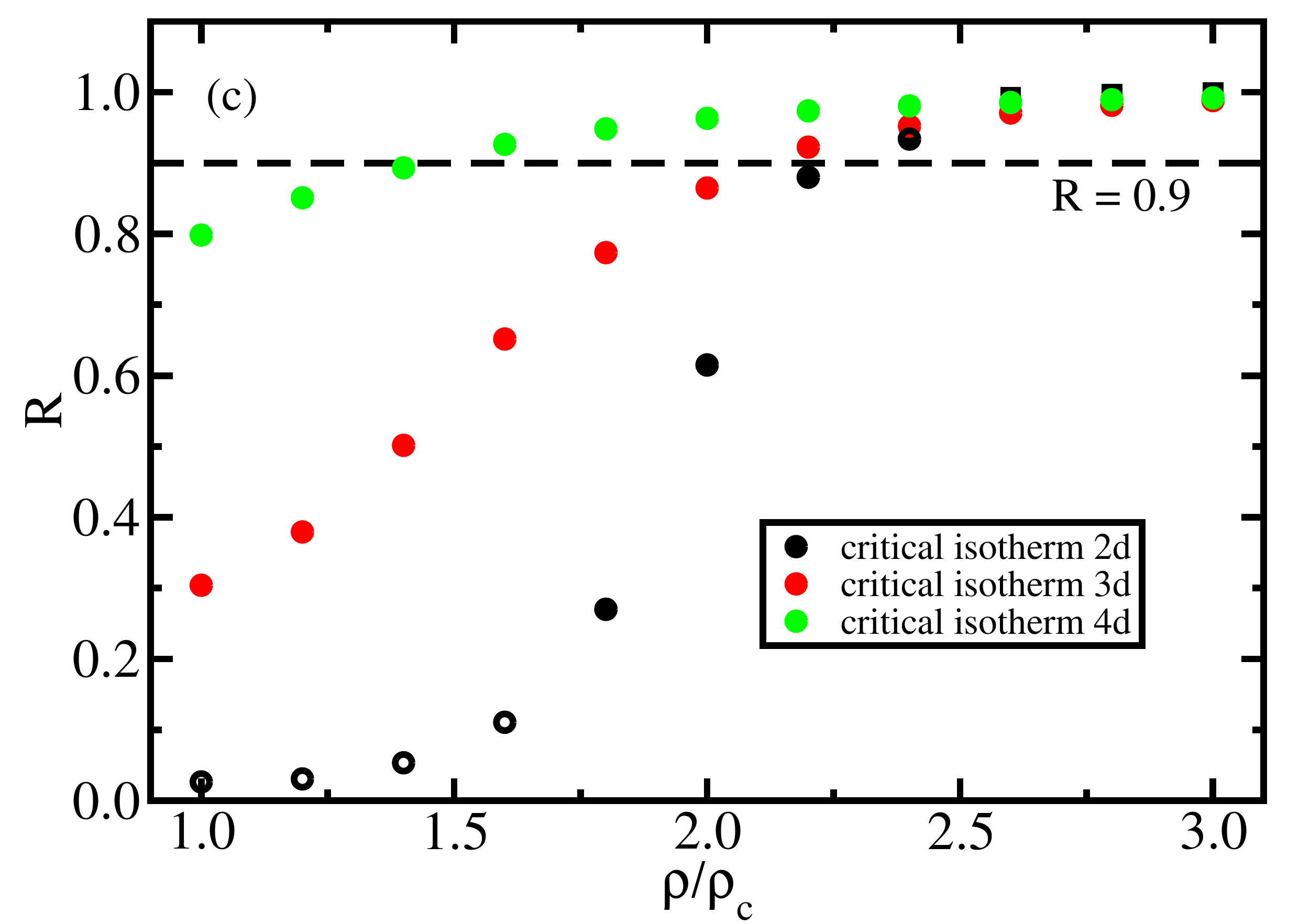} 
	\end{center}
	\caption{The LJ fluid's virial potential-energy correlation coefficient in 2, 3, and 4 dimensions. 
		(a) shows a sketch of the temperature-density phase diagram in which both variables are normalized to their values at the critical point, $T_c$ and $\rho_c$ \cite{smi91,pot98,hlo99}. The black symbols and full curves represent the phase limits of the LJ system in 3d (see, e.g., Refs. \onlinecite{hey15a,cos16} and their references). The orange crosses mark the state points simulated in 2d, 3d, and 4d, whose virial potential-energy correlation coefficients are reported in (b) and (c), the green symbols indicate the three isomorphic state points simulated in 4d (Fig. \ref{4disom_fig}).
		(b) The virial potential-energy correlation coefficient $R$ (\eq{R}) along the critical isochore. There is generally a convergence to the hidden-scale-invariance property characterizing R simple systems defined by $R>0.9$ (dashed horizontal line), a situation that is reached much earlier in four than in three dimensions, where it is reached much earlier than in two dimensions. In two dimensions the system developed ``holes'' close to the critical point (see the main text), which is indicated by the two open symbols.
		(c) The virial potential-energy correlation coefficient along the critical isotherm. There is convergence to the hidden-scale-invariance scenario characterizing R simple systems, a situation that is reached much earlier in four than in three dimensions, where it is reached much earlier than in two dimensions. The two-dimensional system crystallized at the highest densities ($\rho/\rho_c>2.5$), which is indicated by black square symbols; as in (b) the four open symbols at lower densities indicate that the sample developed ``holes'' close to the critical point. 
 		\label{R_fig}}
\end{figure}

In order to systematically compare what happens in different dimensions we studied the variation of the virial potential-energy correlation coefficient $R$ of \eq{R} as one moves away from the critical point along the critical isochore and isotherm, respectively.  In units of $\varepsilon/k_B$ for temperature and $1/\sigma^d$ for density the critical point is given by $(\rho,T)=(0.355,0.515)$ in two dimensions \cite{smi91}, by $(\rho,T)=(0.316,1.312)$ in three dimensions \cite{pot98}, and by $(\rho,T)=(0.34,3.404)$ in four dimensions \cite{hlo99} (the 2d and 3d critical point data were calculated by Monte Carlo (MC) simulations with the LJ  potential truncated at the half box length, the 4d critical point was determined by MC simulations with the potential truncated at $2.5\sigma$). 

Figure \ref{R_fig}(a) gives an overview of the density-temperature thermodynamic phase diagram in which both variables in the standard van der Waals way have been normalized to unity at the critical point. The full black curves indicate the freezing and melting lines for the 3d case, and the orange crosses mark the state points simulated. The results for $R$ are shown in \fig{R_fig}(b) for the critical isochore and in \fig{R_fig}(c) for the critical isotherm (in both figures the horizontal dashed lines mark the (a bit arbitrary) threshold $R=0.9$ defining R simple systems \cite{I}). In two dimensions the system developed visible ``holes'' close to the critical point deriving from large density fluctuations \cite{rov90}; the corresponding simulations are marked by open (black) symbols. In all cases, along both the isochore and the isotherm the correlations increase significantly as one moves away from the critical point. Note that in four dimensions $R$ is fairly large already at the critical point. 

When contemplating these findings one should keep in mind that $R$ is close to unity for the LJ system in three dimensions in the ``ordinary'' condensed liquid phase not too far from the melting line. Our conclusions based on \fig{R_fig} are: 1) The simulations confirm the prediction of Maimbourg and Kurchan that all systems in their condensed-matter (``hard'') regime have strong correlations in high dimensions. 2) There is a striking difference between two, three, and four dimensions, and already in four dimensions the correlations are strong whenever density and temperature are above their critical values.

Before proceeding to discuss the implications of these findings for liquid-state theory we take the opportunity  to demonstrate the existence of isomorphs in four dimensions. The most general method for mapping out an isomorph in the thermodynamic phase diagram makes use of the fact that isomorphs are configurational adiabats \cite{IV,sch14} in conjunction with the following standard fluctuation identity \cite{IV} (in which $\Sex$ is the entropy minus that of an ideal gas at the same density and temperature):

\be\label{isom}
\left(\frac{\partial\ln T}{\partial\ln\rho}\right)_{\Sex}
\,=\,\frac{\langle\Delta W\Delta U\rangle}{\langle(\Delta U)^2\rangle}\,.
\ee
We changed density in steps of 1\%, 2\%, and 5\%, respectively, in each step calculating from \eq{isom} the temperature change needed to keep $\Sex$ constant. An alternative way of generating isomorphic state points, which is limited to LJ-type systems, utilizes the fact that due to invariance of the structure in reduced units, the quantity $h(\rho)/T$ is isomorph invariant where $h(\rho)=A\rho^{12/d}-B\rho^{6/d}$ (the two constants $A$ and $B$, which are (slightly) isomorph dependent, are determined from simulations at a reference state point specifying the isomorph in question; see Refs. \onlinecite{cos16,boh14} for justification and more details of this procedure). Figure \ref{4disom_fig}(a) demonstrates consistency between the two different ways of generating an isomorph in 4d, although for the largest density step (5\%) there is a small disagreement. 

The isomorph invariance of $h(\rho)/T$ signals a breakdown of the theory at low densities at which the above expression for $h(\rho)$ becomes negative. This means that along any isomorph the virial potential-energy correlations must eventually weaken at low densities, which is also observed \cite{I,bai13}. Since $R\rightarrow 1$ in high dimensions for the state points not too far away from the melting line \cite{mai16}, one may speculate that in the $d\rightarrow\infty$ limit there is a phase transition between a phase of increasingly perfect hidden scale invariance and one of poor virial potential-energy correlations \cite{mai16,cos16a}.

Starting from the 4d state point $(\rho\sigma^4,k_BT/\varepsilon)=(0.9,1.0)$ two isomorphic state points were found. The first one $(\rho\sigma^4,k_BT/\varepsilon)=(0.945,1.23)$ was identified using \eq{isom} as described above, the second one $(\rho\sigma^4,k_BT/\varepsilon)=(1.5,6.56)$ was determined using the isomorph invariance of $h(\rho)/T$. Figure \ref{4disom_fig}(b) shows the pair distribution function as a function of the reduced radius for the three state points. The collapse validates structural invariance along the 4d isomorph. Figure \ref{4disom_fig}(c) shows the mean-square displacement as a function of time in reduced units for the same three state points, demonstrating isomorph invariance also of the dynamics.

\begin{figure}
	\begin{center}
		\includegraphics[width=7cm]{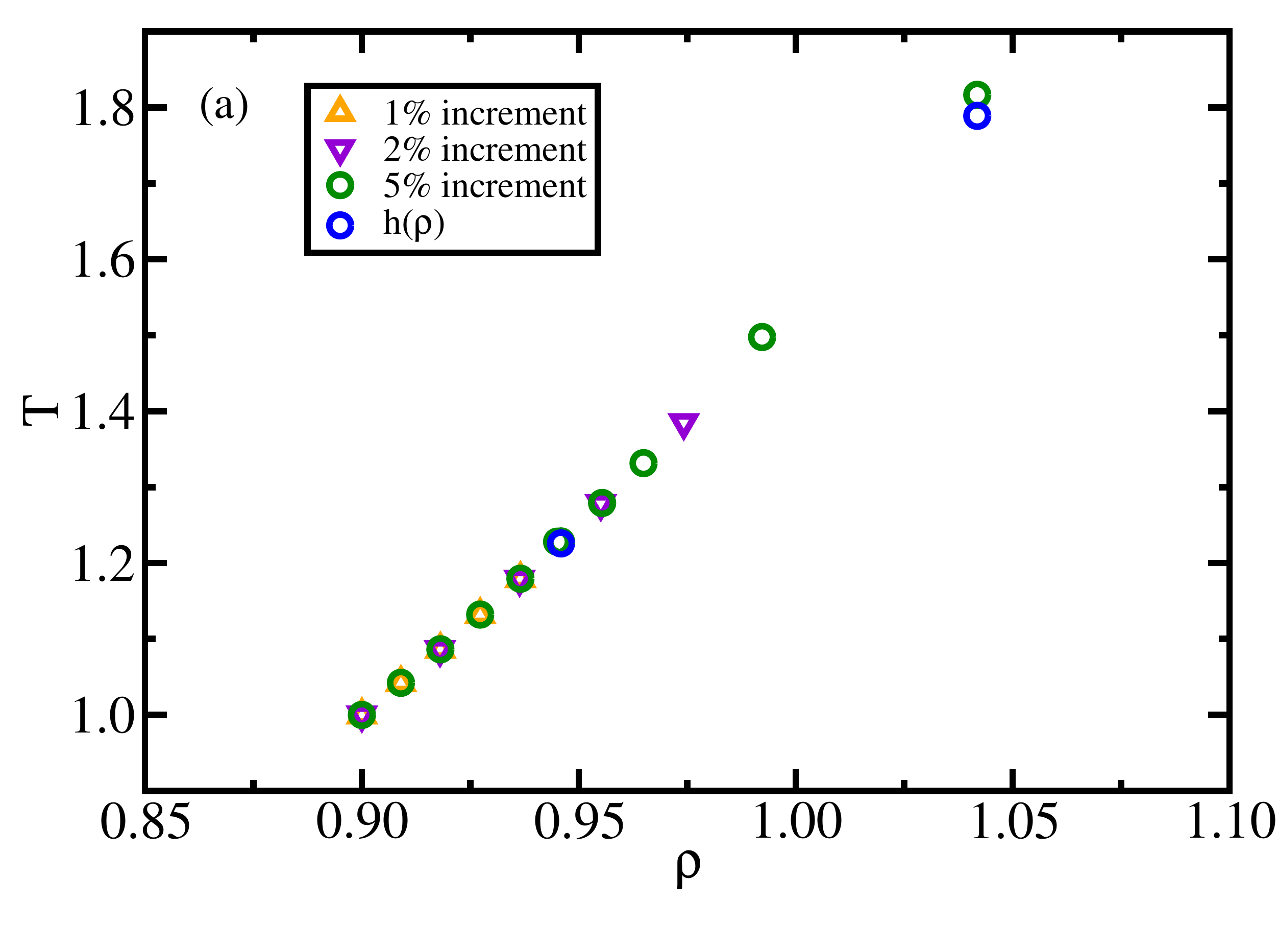} 
		\includegraphics[width=7cm]{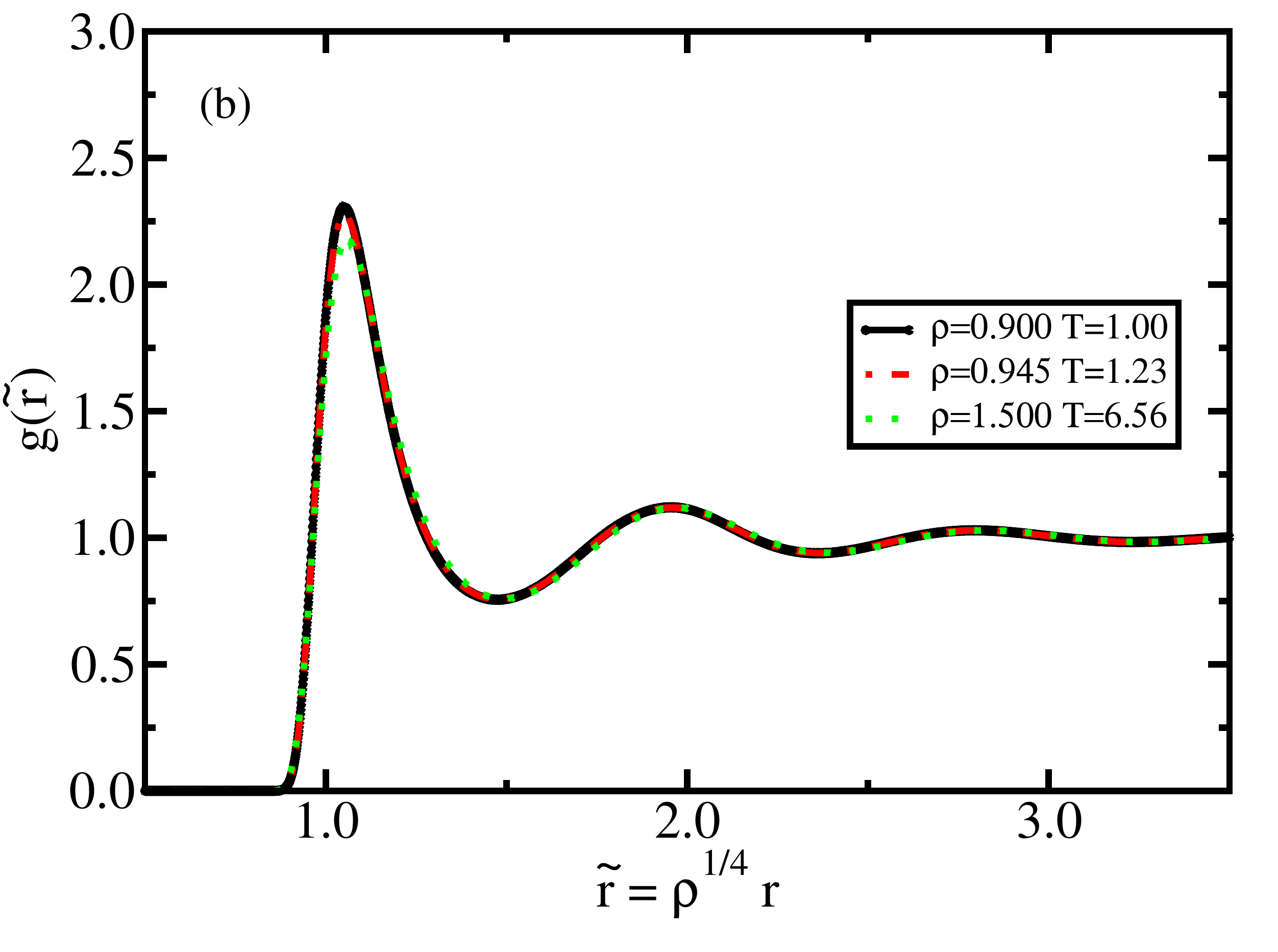} 
		\includegraphics[width=7cm]{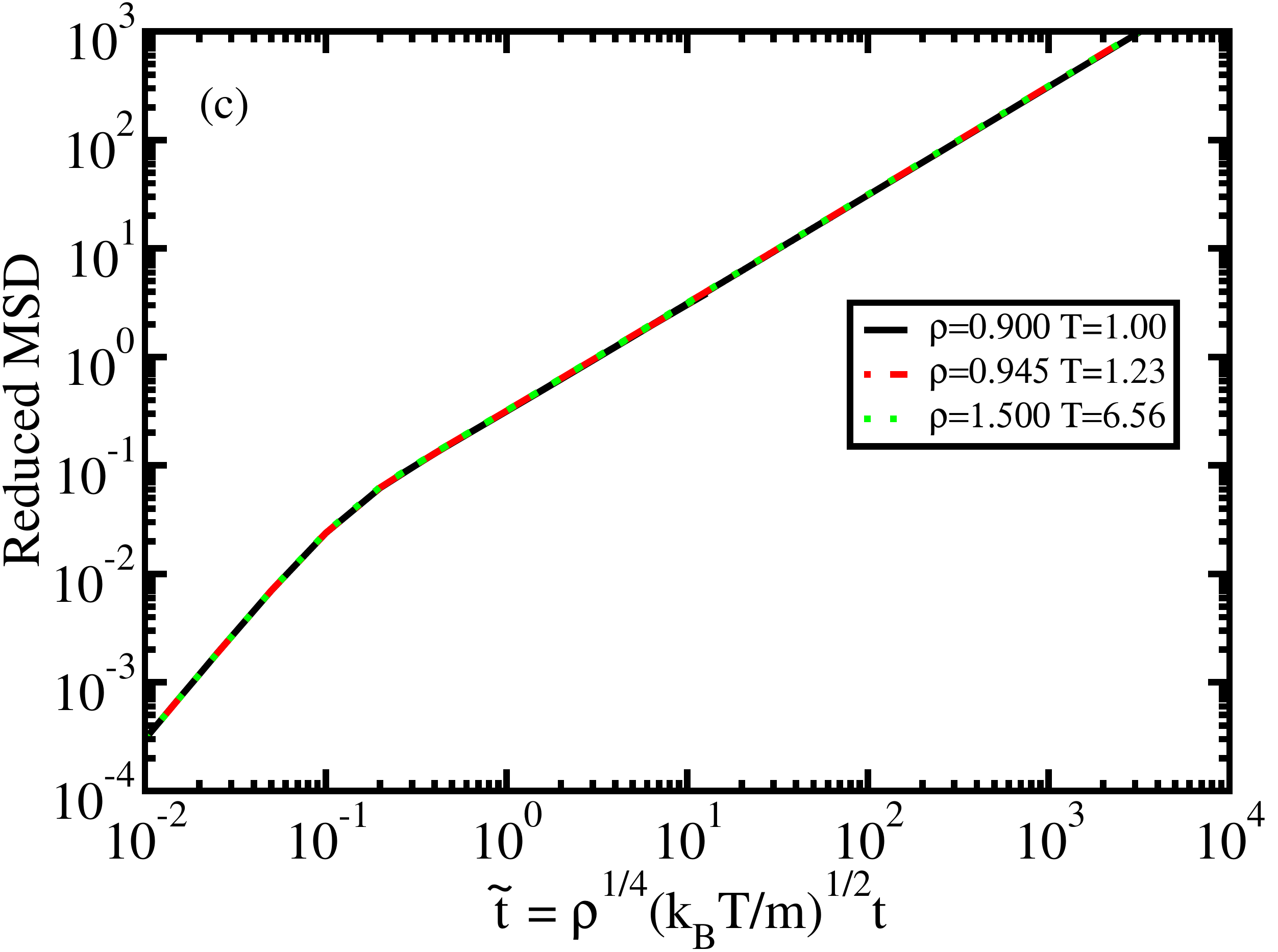} 	
	\end{center}
	\caption{Validation of isomorph invariance in four dimensions based on simulations at three state points on the same isomorph. Starting from the reference state point $(\rho\sigma^4,k_BT/\varepsilon)=(0.9,1.0)$ two isomorphic state points were generated as described in the text. 
	(a) Consistency check of the two different ways of generating isomorphs detailed in the text. The yellow, purple, and green symbols give the results of using \eq{isom} repeatedly for, respectively, a 1\%, 2\%, and 5\% density increase starting from the reference state point; the blue point was calculated by the $h(\rho)$ method described in the text.
	(b) The pair distribution function at the three isomorphic state points plotted as a function of the reduced pair distance. The collapse demonstrates structural invariance along the isomorph.
	(c) Reduced mean-square displacement as a function of reduced time for the same three isomorphic state points, demonstrating isomorph invariance of the dynamics (reduced units are defined in Ref. \onlinecite{IV}).
	\label{4disom_fig}}
\end{figure}

Turning back to the dimensionality dependence of the virial potential-energy correlations, our findings may be summarized as follows. Above the critical point as the number of dimensions increases the LJ system converges rapidly to the state of perfect hidden scale invariance shown by Maimbourg and Kurchan to characterize the high-dimensional limit. Assuming that the challenge of generalizing arbitrary systems to arbitrary dimensions has been addressed, we conjecture the following: 1) All systems (also molecular ones) obey hidden scale invariance in sufficiently high dimensions in their condensed phases; 2) the rate with which this property translates into lower dimensions depends on the system in question. In other words, if one defines the van der Waals scaled density $\tilde{\rho}\equiv \rho/\rho_c$ and temperature $\tilde{T}\equiv T/T_c$, we conjecture that $R(\tilde{\rho},\tilde{T})\rightarrow 1$ as $d\rightarrow\infty$ for all systems, at least whenever $\tilde\rho>1$ and $\tilde T>1$. The rate of convergence determines whether or not the system is R simple in three dimensions.

If the above conjecture is correct, any system at any given condensed-matter state point has a ``transition region'' of dimensionalities above which it becomes R simple. This range of dimensions is located below three dimensions for systems that are R simple in three dimensions (at the state point in question) and above three for those that are not.

An important task for the future will be to construct a systematic $1/d$ expansion taking one from the case of guaranteed R simple behavior as $d\rightarrow\infty$ to three dimensions. Hints of how this may be done were given in Ref. \onlinecite{cha14} for the HS case, but a more general approach is needed. We find it conceivable that future textbooks in liquid-state theory start by deriving a simple and general theory for the limit of high dimensions and subsequently translates this into three dimensions via a $1/d$ expansion, but clearly much remains to be done before this becomes reality.

\acknowledgments{We are indebted to Thibaud Maimbourg for his comments to an early draft of this paper. This work was supported in part by the Danish National Research Foundation via grant DNRF61.}

\end{document}